\documentclass[aps,superscriptaddress,twocolumn,]{revtex4}
\usepackage{bm}
\usepackage{epsfig}
\usepackage{times}
\usepackage{bbm}
\usepackage{amssymb}
\usepackage{amsmath}
\usepackage{comment}
\usepackage{natbib}
\usepackage{color}
\usepackage{graphicx}
\usepackage[normalem]{ulem} 
\usepackage[latin1]{inputenc}
\usepackage[loose,nice]{units}       

\begin{document}
\title{Absorbers as detectors for unbound quantum systems}

\author{S{\o}lve Selst{\o}}
\affiliation{Faculty of Technology, Art and Design, Oslo Metropolitan University, NO-0130 Oslo, Norway}

\begin{abstract}
Complex absorbing potentials are frequently imposed when simulating unbound quantum systems. While this is usually done solely in order to avoid artifacts at the numerical boundary, we show how absorbers may also be used to probe the characteristics of the particle undergoing absorption. This way, information about the removed particles may be retained. Moreover, we explain how -- and when -- such absorbers act as detectors in the sense that they allow for simulating the act of measurement in dynamical simulations.
\end{abstract}

\maketitle

\section{Introduction}
\label{IntroductionSeq}

As unbound quantum systems may extend arbitrarily far, their numerical description tend to be quite demanding -- if not to say unfeasible -- in many cases. On order to simulate unbound quantum system on a numerical domain smaller than the actual extension of the wave function, absorbing boundary conditions are frequently imposed. This allows for outgoing waves to be attenuated without imposing artifacts such as unphysical reflections at the boundary.

One very common way to impose absorbing boundary conditions is to augment the Hamiltonian with a {\it Complex Absorbing Potential} (a CAP)~\cite{Weisskopf1930, Kosloff1986, Riss1993,Muga2004}. When CAPs are imposed, it is typically with the sole purpose of removing outgoing waves -- with less regard for what is removed.
However, CAPs may, in many cases, be interpreted as a models for detectors~\cite{Kosloff1986,Kvaal2011}.
In particular, this way of using CAPs has proven useful when calculating arrival times in quantum physics~\cite{Allcock1969}. In such a context, arrival times may be determined from the loss in the norm of the wave function, the {\it absorption rate}, that the CAP induces. In the case of too strong absorbing potentials, such rates are, in fact, suppressed. This, in turn, is a manifestation of the quantum Zeno effect~\cite{Echanobe2008}.

Regardless of wether CAPs are used merely as means to truncate the numerical domain or in order to model detection, they are, almost without exception, imposed on heuristic or pragmatic grounds, and their specific shape are usually chosen solely in order to induce as little artificial reflection as possible \cite{Manolopoulos2002}. Refs.~\cite{Halliwell1999,Ruschhaupt2004} are exceptions to this rule, however. In both these works CAPs are derived from more fundamental principles. In the latter case, this is done under very specific conditions,
while the former, on the contrary, features a very generic approach, an approach which involves a two-level detector which is coupled to an environment in order to encompass the irreversible nature of measurement. This, in turn, underlines that detection really is a Markovian process, and the proper framework is, in fact, that of open quantum systems.

Models such as the ones addressed above succeed, in certain respects, in describing how detection influences the quantum system in question. However, the main purpose of detection is, after all, extracting physical information of the quantum system in question. Detector models which incorporate this aspect are scarce. This may seem odd; since we, in fact, know precisely what the CAP removes at each time, this information may be retain and aggregated.
In fact, for one-particle systems this is quite easy, bordering on the trivial. While somewhat less trivial, it may, as explained in \cite{Selsto2021}, be generalized to any number of particles exposed to a CAP which acts as a one-particle operator.

In the present work we will study how various ways of implementing CAPs affect the coherence properties of a quantum system, {\it and} we will use the CAP to {\it probe} the outgoing, attenuated waves so that the information we remove is retained. This way we will calculate both angular and energy/momentum distributions of unbound quantum particles -- quantities of experimental interest. We will see that the interpretation of a CAP as a detector model only makes sense under certain conditions. Specifically, a CAP acts as a detector when we analyse the absorbed part, in terms of projective measurements, in a basis in which the CAP is diagonal. If we, on the other hand, apply a non-diagonal absorber, coherence properties are maintained, thus allowing for extracting physical results which are very robust indeed in the sense that they are virtually independent of the CAP.

In the next section the theory is outlined. Contrary to Ref.~\cite{Selsto2021}, we will here be concerned with the one-particle case. Numerical examples are presented and discussed in Sec.~\ref{ResultsSeq}. The generalization to the many-particle case is outlined towards the end of this section while conclusions are drawn in Sec.~\ref{ConclusionsSeq}.

\section{Theory}
\label{TheorySeq}

We will briefly explain how a CAP causes depletion in the wave function of a quantum system -- and how the CAP may be used to probe the absorbed part in order to retain the desired information.

\subsection{Wave function depletion}
\label{WFdepletionSeq}

The evolution of the wave function for a quantum system
is dictated by the Schr{\"o}dinger equation,
\begin{equation}\label{TDSE}
i \hbar \frac{\mathrm{d}}{\mathrm{d}t} \Psi = H \Psi,
\end{equation}
or generalizations thereof. The Hamiltonian $H$, which is Hermitian, is the operator corresponding to the total energy of the system. Absorbing boundary conditions may be imposed by replacing the Hamiltonian $H$ in Eq.~(\ref{TDSE}) with an effective non-Hermitian Hamiltonian
\begin{equation}
\label{AugmentHwithCAP}
H_\mathrm{eff} = H - i \Gamma ,
\end{equation}
where the CAP, $\Gamma$, may be a local potential, i.e., a potential which depends on position only, or non-local. In any case, it should be chosen such that it only affects the dynamics near the boundary of the numerical domain and leaves the interior region unaffected.
Here we will also insist that the CAP is Hermitian,
\begin{equation}
\label{HermitianCAP}
\Gamma^\dagger = \Gamma .
\end{equation}
Moreover, it should be positive semi-definite,
\begin{equation}\label{PisitivSemiDef}
  \Gamma \geq 0 \quad .
\end{equation}
This ensures that the norm of the corresponding wave function cannot increase, which would be unphysical.
A {\it decrease} in norm, on the other hand, is physical in the sense that it corresponds to a depletion of the wave function due to quantum particles leaving the numerical domain.


In a many-particle context, this is devastating if you wish to simulate the remainder of a multi-particle quantum system after the full absorption of one particle; the wave function will, according to Eq.~(\ref{TDSE}) with the effective Hamiltonian of Eq.~(\ref{AugmentHwithCAP}), vanish completely. This may, however, be remedied by realizing that absorption is a Markovian process; there is no memory of any particle which is absorbed.
Correspondingly, the GKLS equation, see \cite{Lindblad1976, Gorini1976}, emerges as the natural candidate for generalizing the Schr{\"o}dinger equation into one that is able to resolve dynamics with a decreasing number of particles.
In \cite{Selsto2010} it is detailed how this generalization comes about.
And in \cite{Kvaal2011} this formulation is adapted to a Multi-Configuration Time-Dependent Hartree Fock formulation.


In the following we will, however, consider the evolution as dictated by the Schr{\"o}dinger equation,
Eq.~(\ref{TDSE}),
which is makes sense when we restrict ourselves to the one-particle case.
From time $t$ to $t + \Delta t$ the wave function evolves from $\Psi(t)$ to
\begin{equation}\label{WFevolve}
\Psi(t + \Delta t) = \Psi(t) - \frac{i}{\hbar} H \Psi(t) \Delta t - \frac{1}{\hbar} \Gamma \Psi(t) \Delta t + \mathcal{O}(\Delta t^2) .
\end{equation}
The CAP
causes a part of the wave function,  $\Delta t/ \hbar \; \Gamma \Psi(t)$, to be removed.
This removed part may be analysed on the fly and added to the total probability distribution of interest. However, in arriving at the proper way of doing so,
we find it more instructional to formulate the problem in terms of a density matrix, $\rho(t) = | \Psi \rangle \langle \Psi |$, instead. For $\rho$, the expression analogous to Eq.~(\ref{TDSE}) reads
\begin{equation}\label{DensMatEvolve}
\rho(t+ \Delta t) = \rho(t) - \frac{i}{\hbar} [H, \rho] \Delta t - \frac{1}{\hbar} \{ \Gamma, \rho \} \Delta t  + \mathcal{O}(\Delta t^2) .
\end{equation}
The part which has been removed from the density matrix during this time step, is given by the anti-commutator above.

\subsection{The momentum/energy differential spectrum}
\label{EnergyDistSeq}

For a single-particle system, the anti-commutator in Eq.~(\ref{DensMatEvolve}) is an effective one-particle density matrix which can be analyzed:
\begin{equation}
\label{SigmaDef}
d \sigma = \frac{1}{\hbar} \{ \Gamma, \rho \} \, dt .
\end{equation}
Suppose now that we are interested in obtaining the probability distribution differential in momentum for the unbound particle, $\partial P/ \partial p$. This can be calculated by aggregating the elements diagonal in momentum eigen states $| p \rangle$ of $\Delta t/\hbar \; \{ \Gamma, \rho \}$:
\begin{equation}\label{dPdEtillegg}
\frac{\mathrm{d}}{\mathrm{d} t}  \frac{\partial P}{\partial p} =
\frac{1}{\hbar} \langle p | \{\Gamma, \rho \} | p \rangle  ,
\end{equation}
or, for those more inclined towards formulations in terms of Projection-Valued Measures (PVMs):
\begin{equation}\label{dPdEtilleggPVM}
\frac{\partial P}{\partial p} =
\int_{t=0}^{t=\infty} \, \mathrm{Tr} \, \left( | p \rangle \langle p | d \sigma  \right).
\end{equation}

%
%

For the CAP, the simplest and, arguably, most natural choice is that of a local potential, i.e., one that is
diagonal in position $x$:
\begin{equation}\label{CAPx}
\Gamma_x = \int dx \; \gamma(x) | x \rangle \langle x | ,
\end{equation}
where $|x \rangle$ are position eigen states and the CAP position function $\gamma(x)$ is zero in the interior of the domain and positive near the boundary of the numerical domain.
Here `$\int dx$' is to be taken as the definite integral over all space, in all dimensions. According to Eq.~(\ref{dPdEtillegg}), the momentum-differential probability distribution of the unbound part of the wave function may be determined as
\begin{align}
\nonumber
&
\frac{\partial P}{\partial p} =
\frac{1}{\hbar} \int_0^\infty dt \; \langle p | \{ \Gamma_x, \rho(t) \} | p \rangle =
\\
\nonumber
&
\frac{1}{\hbar} \int_0^\infty dt \; \langle p | \int dx' \gamma(x') \; \left\{  | x' \rangle \langle x' | , |\Psi(t) \rangle \langle \Psi(t) | \right\} | p \rangle = \\
\label{dPdPfromCAPx}
&
\frac{2}{\hbar} \mathrm{Re} \int_0^\infty dt \; \Phi(p; t) \mathcal{F} \{ \gamma(x) \Psi^*(x; t)\} (p)
\end{align}
where $\mathcal{F}$ is the Fourier transform and $\Phi(p; t)=\mathcal{F}\{\Psi(x; t)\}(p)$ is the momentum wave function.
While the artificial CAP may influence the physical system, Eq.~(\ref{dPdOmegaCAPx}) should, at least, produce the actual asymptotic momentum distribution of the quantum particle in
the limit $\gamma \rightarrow 0^+$.
Note that Eq.~(\ref{dPdPfromCAPx}) provides a {\it coherent} sum of contributions emerging at different times.

This differs from the situation we would have when the CAP is diagonal in momentum or energy rather than position. Such a CAP would not be given by any local potential. Examples of implementations of such non-local CAPs seen in literature could be, e.g., {\it the Transformative CAP} \cite{Riss_1998}, {\it the Reflection-Free CAP} \cite{Moiseyev_1998} or {\it Infinite Range Exterior Complex Scaling} \cite{Scrinzi2010}.
In our context, we insist that the CAP remains Hermitian, cf. Eq.~(\ref{HermitianCAP}); otherwise, there would be Hermitian contributions to the effective Hamiltonian, Eq.~(\ref{AugmentHwithCAP}), in the CAP region which would alter the physics of the particle undergoing absorption.
We also insist that the CAP is written in terms of projections as in Eq.~(\ref{CAPx}). However, a straight forward replacement of the position $x$ with the momentum $p$ in Eq.~(\ref{CAPx}) is not adequate as such an implementation would impose absorption throughout the numerical domain, not just in the vicinity of the boundary. Instead we propose the following energy absorber:
\begin{equation}\label{CAPe}
  \Gamma_\varepsilon = \int d\varepsilon  \; \mu(\varepsilon) | \varphi_\varepsilon \rangle \langle \varphi_\varepsilon | ,
\end{equation}
where the $|\varphi_\varepsilon \rangle$-s are energy normalized eigen functions of the Hamiltonian
\begin{subequations}\label{ModifiedHam}
\begin{align}
\label{ModifiedHamA}
&
H_\varepsilon = H + V_\varepsilon(x) \quad \text{where} \\
\label{ModifiedHamB}
&
V_\varepsilon(x) = \left\{
\begin{array}{ll}
\infty , & x \in D_\mathrm{I} \\
0 , & x \notin D_\mathrm{I}
\end{array}
\right. .
\end{align}
\end{subequations}
Here, $D_\mathrm{I}$ is the interior of the numerical domain; typically it is given by all positions $x$ which are such that $|x|\leq R$ for some finite distance $R$ from the origin. With this the eigen functions $\varphi_p(x) = \langle x | \varphi_p \rangle$ are only supported for $x \notin D_\mathrm{I}$. Correspondingly, the CAP of Eq.~(\ref{CAPe}) is both energy and position dependent, contrary to the strictly position-dependent CAP of Eq.~(\ref{CAPx}). The CAP function $\mu(\varepsilon)$, however, is purely energy dependent. As with $\gamma(x)$ in Eq.~(\ref{CAPx}), it should be
positive.

In obtaining the momentum or energy distribution using the CAP of Eq.~(\ref{CAPe}), we could, once again, calculate the diagonal elements of Eq.~(\ref{dPdEtillegg}) in the momentum basis, $\{ | p \rangle \}$.
However,
energy spectra are calculated more conveniently using the $| \varphi_\varepsilon \rangle$ basis, in which the CAP is diagonal.
This is admissible since these energy eigen states form a complete set within the domain in which absorption takes place -- with the energy eigen values of $H_\varepsilon$ and the actual (Hermitian) Hamiltonian $H$ coinciding. Moreover, as we will see, this allows us to use the CAP of Eq.~(\ref{CAPe}) to model energy-measurement.
We arrive at the following energy distribution:
\begin{align}
& \nonumber
\frac{\partial P}{\partial \varepsilon} =
\int_{t=0}^{t=\infty} \mathrm{Tr} \left( | \varphi_\varepsilon \rangle \langle \varphi_\varepsilon | d\sigma_\varepsilon \right)
=
\\ \nonumber
&
\frac{1}{\hbar} \int_0^\infty dt  \; \langle \varphi_\varepsilon | \{ \Gamma_\varepsilon , \rho(t) \} | \varphi_\varepsilon \rangle = \\
&
\nonumber
\frac{1}{\hbar} \int_0^\infty dt  \; \langle \varphi_\varepsilon | \int d \varepsilon' \mu(\varepsilon')
\{ | \varphi_{\varepsilon'} \rangle \langle \varphi_{\varepsilon'}  | , |\Psi(t) \rangle \langle \Psi(t) | \} | \varphi_\varepsilon \rangle = \\
&
\label{dPdPfromCAPe}
\frac{2}{\hbar} \int_0^\infty dt \; \mu(\varepsilon) |\langle \varphi_\varepsilon | \Psi (t) \rangle |^2 .
\end{align}
We note that, contrary to Eq.~(\ref{dPdPfromCAPx}), this integral is {\it incoherent} and manifestly positive. Again, the correct, asymptotic distribution should be obtained by extrapolating the strength of the CAP function, in this case $\mu(\varepsilon)$, to zero. However, as we will apparent in Sec.~\ref{ResultsSeq}, it may also be interesting to study the spectra obtained with finite $\mu(\varepsilon)$.

\subsection{Angular distribution}
\label{AngularDistSeq}

In addition to energy/momentum differential distributions, distributions differential in position, or rather, {\it direction} are also of interest experimentally.
For a three-dimensional system we may write the position eigen state $|x \rangle$ in terms of spherical coordinates as $|r, \Omega \rangle$ and introduce the additional assumption on the CAP of Eq.~(\ref{CAPx}) that it is isotropic:
\begin{equation}
\label{CAPxIsotropic}
\Gamma_r = \int r^2 dr d \Omega \; \gamma(r) | r, \Omega \rangle \langle r, \Omega | .
\end{equation}
As in the case of Eq.~(\ref{dPdPfromCAPe}), the position distribution obtained from the position diagonal CAP of Eq.~(\ref{CAPxIsotropic}) becomes an incoherent time-integral:
\begin{align}
\label{TriplyDiffPosition}
& \frac{\partial^2 P}{\partial r \partial \Omega} =
\frac{1}{\hbar} \langle r, \Omega | \int_0^\infty dt \; \int r'^{2} dr' d\Omega' \; \gamma(r')
\\ \nonumber &
\times
\left\{ | r', \Omega' \rangle \langle r', \Omega' | , | \Psi \rangle \langle \Psi | \right\} | r, \Omega \rangle =
\\ \nonumber &
\frac{2}{\hbar} \int_0^\infty dt \, \gamma(r) \left| \langle r, \Omega |  \Psi \rangle \right|^2
.
\end{align}
As the distribution in radial distance $r$ is usually less interesting than the distribution in $\Omega$, we integrate out the $r$-dependence and arrive at
\begin{equation}
\label{dPdOmegaCAPx}
\frac{\partial P}{\partial \Omega} = \frac{2}{\hbar} \int_0^\infty dt \; \int_0^\infty r^2 dr \;
\gamma(r) |\Psi(r,\Omega)|^2 .
\end{equation}
The analogous expression in two-dimensional polar coordinates reads
\begin{equation}
\label{dPdThetaCAPx}
\frac{\partial P}{\partial \theta} = \frac{2}{\hbar} \int_0^\infty dt \; \int_0^\infty r dr \;
\gamma(r) |\Psi(r,\theta)|^2 .
\end{equation}
Determining angular distributions in this way lends itself to particularly straight forward implementation with wave functions expressed in spherical or polar coordinates. Moreover, the isotropic, sparse nature of the CAP allows for rather simple implementation when resolving the dynamics of the system.
%
%

\section{Numerical Examples}
\label{ResultsSeq}

In the following we will illustrate the approaches outlined above to absorption and analysis of two particular unbound one-particle systems. They are both rather simple and generic, and they serve well to illustrate the close correspondence between detectors and absorbers.

\subsection{Energy spectra for a one-dimensional system}
\label{Example1Seq}

Our first example is a one-dimensional one. Correspondingly, $x$ and $p$ are scalar quantities in this context. The particle is initially trapped in the ground state of a confining potential, which features a finite number of bound states and a continuum. The confining potential has a Gaussian shape:
\begin{equation}\label{GaussPot}
V(x) = -V_0 \exp\left(-\frac{x^2}{2 \sigma_V^2} \right) .
\end{equation}
The particle is exposed to an explicitly time-dependent perturbation; the total (Hermitian) Hamiltonian of the system reads
\begin{equation}\label{Hamiltonian1D}
H=-\frac{\hbar^2}{2m} \frac{d^2}{dx^2} + V(x) + q E(t) x,
\end{equation}
where $q$ is the charge of the particle and
\begin{subequations}\label{LaserFields}
\begin{align}
\label{DoubleField}
& E(t) = E_\mathrm{Pulse}(t) + E_\mathrm{Pulse}(t-T+\tau)  \quad \text{where} \\
&
\label{SinglePulse}
E_\mathrm{Pulse}(t) = \left\{ \begin{array}{ll} E_0 \sin^2 \left( \frac{\pi}{T} t\right) \sin (\omega t) , & 0 \leq t \leq T \\
0, & \text{otherwise} \end{array} \right. .
\end{align}
\end{subequations}
The system may serve as a model atom exposed to two consecutive pulses of radiation. Because of this perturbation there is a finite probability that the trapped particle is liberated after the interaction, and the outgoing waves are subject to interference.

The time-dependent Schr{\"o}dinger equation, Eq.~(\ref{TDSE}), is solved numerically by means of a split operator technique in which the double spatial derivative of Eq.~(\ref{Hamiltonian1D}) is approximated by a finite difference scheme. It is solved with the effective Hamiltonian of Eq.~(\ref{AugmentHwithCAP}) with the Hermitian part given by Eq.~(\ref{Hamiltonian1D}). In the first case, we let the anti-Hermitian part, the CAP $-i \Gamma$, be a local absorber of the form of Eq.~(\ref{CAPx}). The CAP function reads
\begin{equation}
\label{SquareCAPx}
\gamma(x) = \left\{ \begin{array}{ll} \gamma_0 (x-R)^2, & |x|> R \\
0, & \text{otherwise} \end{array} \right.
\end{equation}

\begin{figure}
\begin{tabular}{c}
\includegraphics[width=7cm]{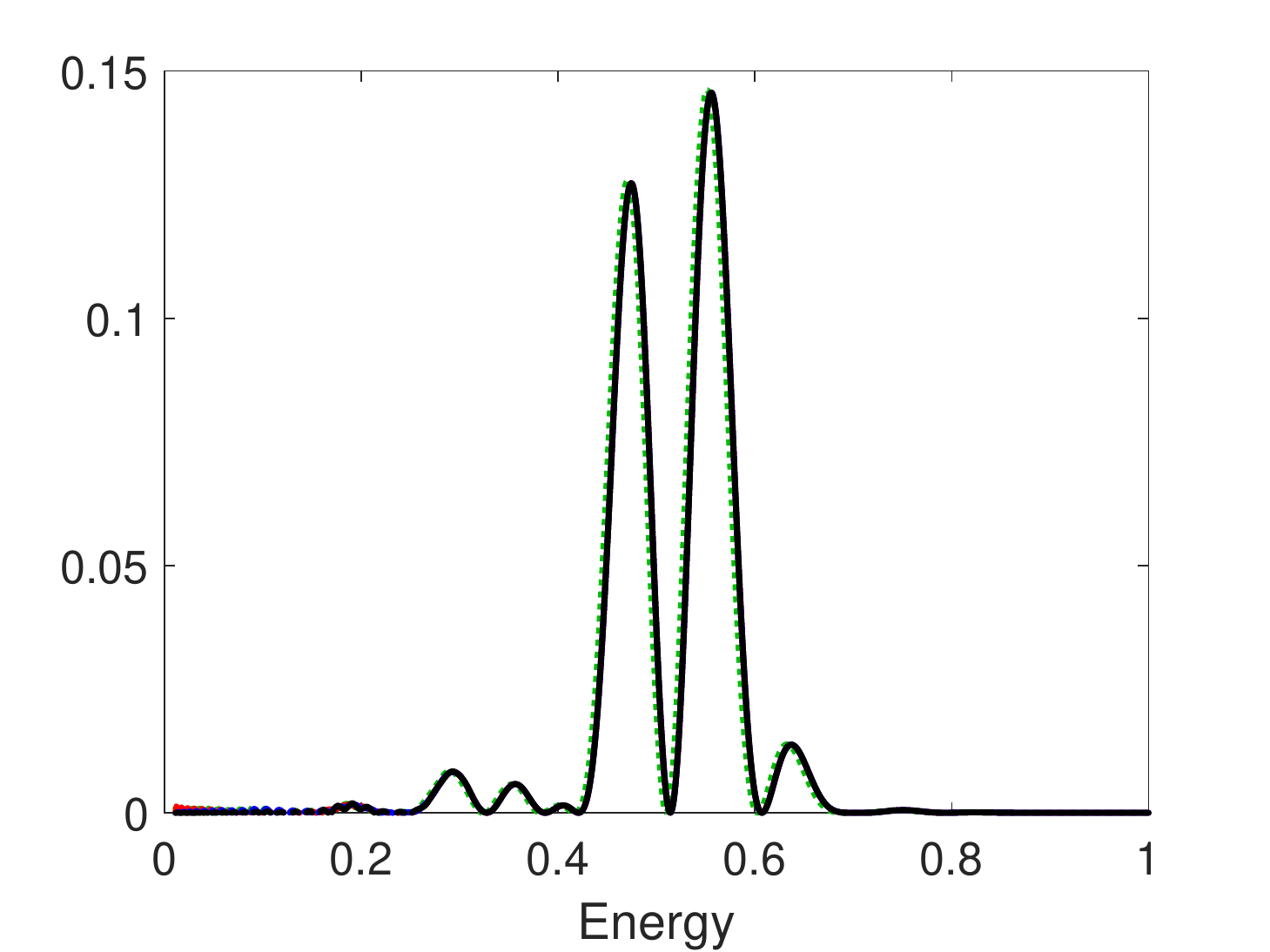} \\
\includegraphics[width=7cm]{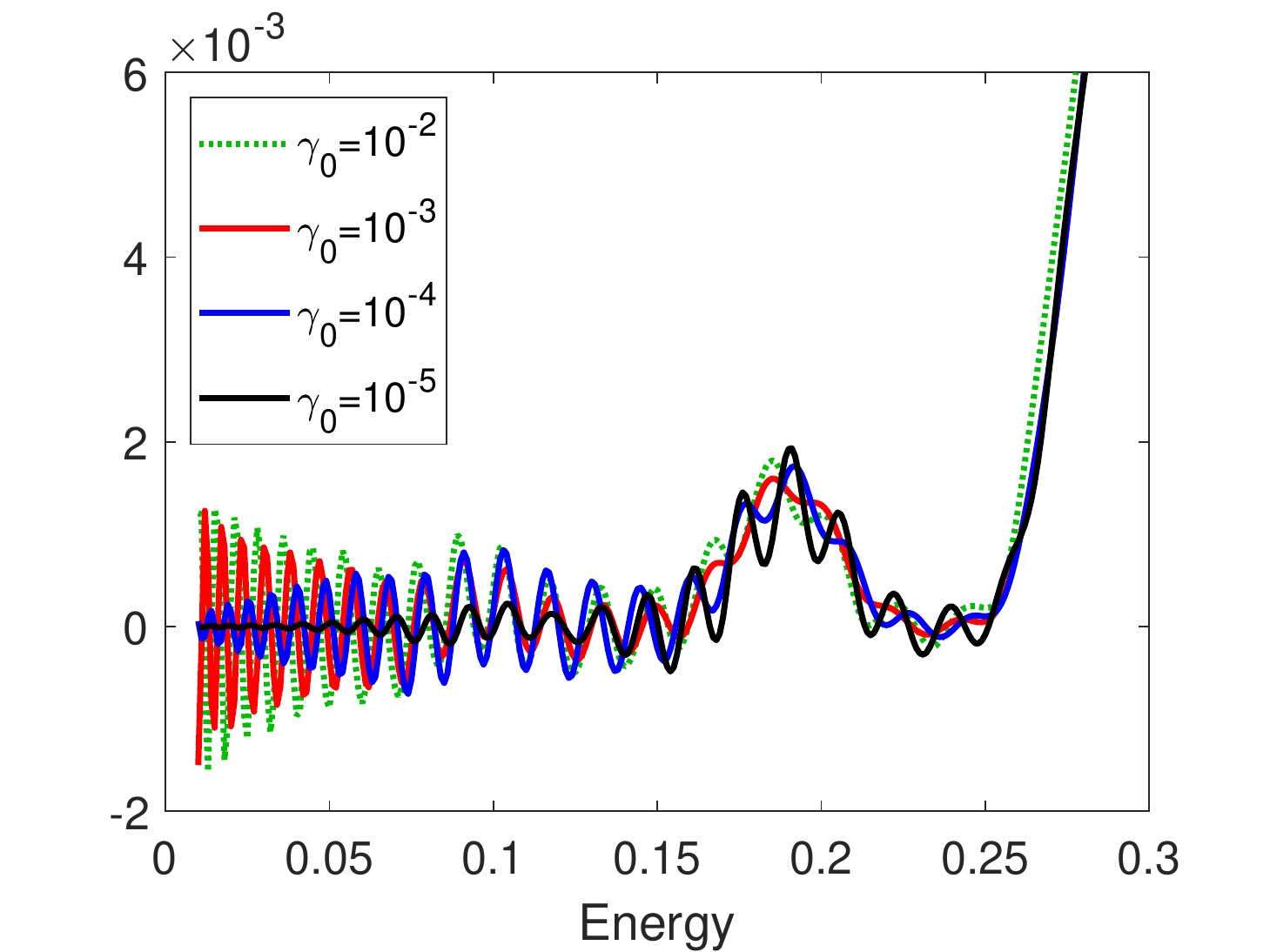} \\
\includegraphics[width=7cm]{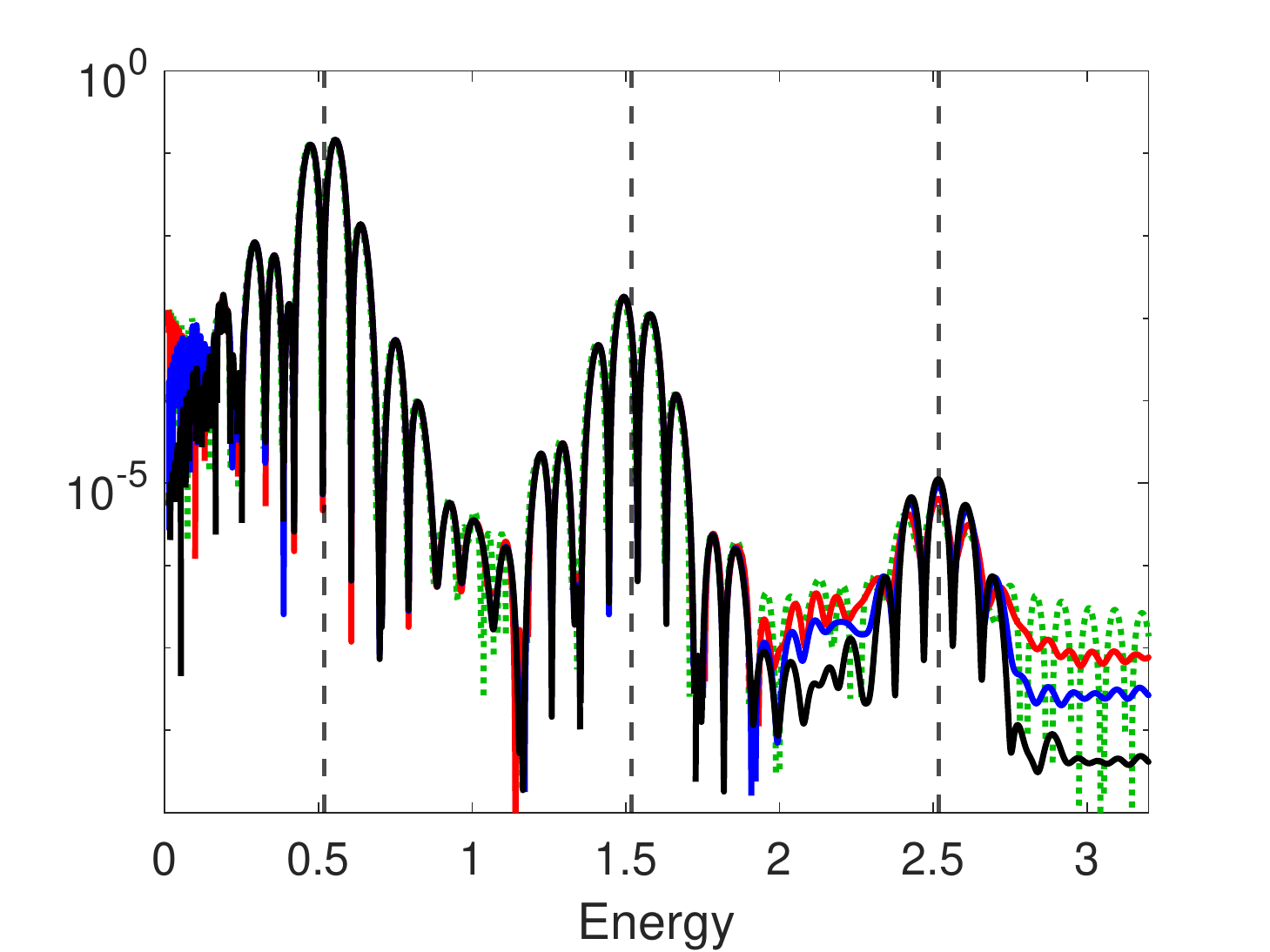} \\
\end{tabular}
\caption{Upper panel: Energy distribution from the unbound part of the wave function obtained from a local CAP. The spectra are plotted for various CAP strengths. Middle panel: A close up in the low energy region. Lower Panel: The energy distributions plotted with a logarithmic $y$-axis.}
\label{dPdEfromCAPxFigure}
\end{figure}

The interference between outgoing wave components liberated at different times causes a rich structure in the emerging energy distribution of the unbound particle.
Figure~\ref{dPdEfromCAPxFigure} shows this distribution calculated by using Eq.~(\ref{dPdPfromCAPx}).
It is plotted against energy rather than momentum; in the CAP region, the confining potential may safely be neglected so that $\varepsilon = p^2/2m$.
We use units defined by choosing $\hbar$, $m$ and $-q$ as the unit of their respective quantities. In these units, the confining potential has the depth $V_0=0.6$ and the width $\sigma_V=3$, cf. Eqs.~(\ref{GaussPot}). The corresponding ground state energy is $-0.48$. The time dependent perturbation is characterized by the strength $E_0=2$, the angular frequency $\omega=1$, and the delay $\tau= 5$ between the pulses, cf. Eqs.~(\ref{LaserFields}). Each pulse has a duration corresponding to ten optical cycles, $T=10 \times 2\pi/ \omega$. The CAP onset is $R=200$~length units.

In Fig.~\ref{dPdEfromCAPxFigure} the emerging spectra are calculated using various absorber strengths $\gamma_0$. It is striking to see that not only does the spectrum converge as the strength of the CAP function decreases; apart from the low energy region, it is virtually independent of the CAP strength.
%
This feature does in no way rely on the specific shape of the CAP function; other choices than the one of Eq.~(\ref{SquareCAPx}) display the same behaviour.
The discrepancies we see at low energies are clearly unphysical, not only because of the $\gamma_0$ dependence, but also because they produce ``negative probabilities'' -- also with a comparatively week absorber. This undesired feature, which diminish with decreasing $\gamma_0$, seems to be related to the fact that hard absorption does induce some artificial reflections.

With the external field of Eqs.~(\ref{LaserFields}) interpreted as an external electric field, the structure seen in the upper panel of Fig.~\ref{dPdEfromCAPxFigure} corresponds to absorption of one photon. With a logarithmic $y$-axis, we also clearly see structures corresponding to absorption of two and three photons as well.
A somewhat stronger $\gamma_0$-dependence is seen at the multi-photon peaks, indicating that decreasing absorption strength is needed in order to resolve peaks corresponding to a higher number of photons.

\begin{figure}
\begin{tabular}{c}
\includegraphics[width=7cm]{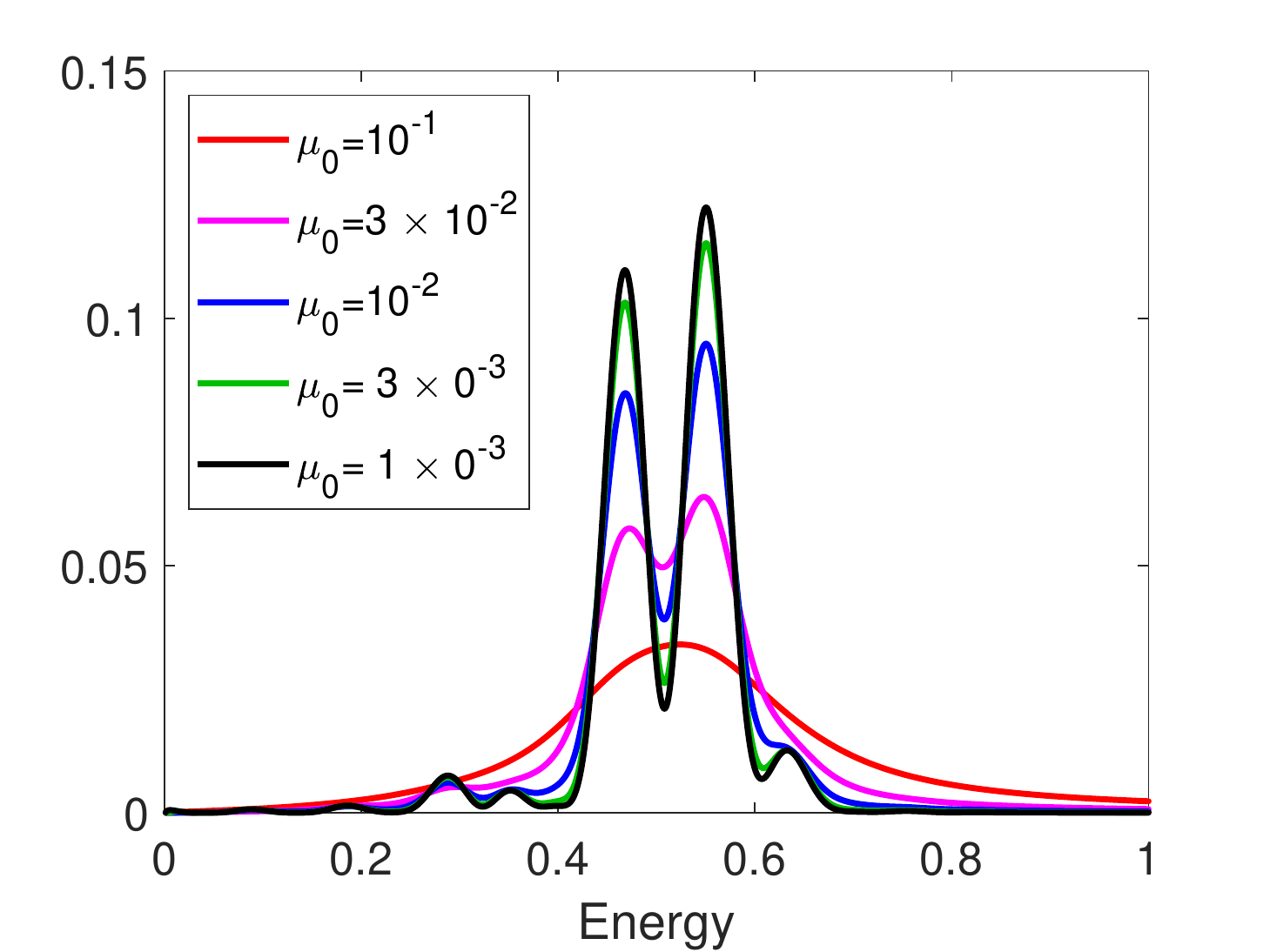} \\
\includegraphics[width=7cm]{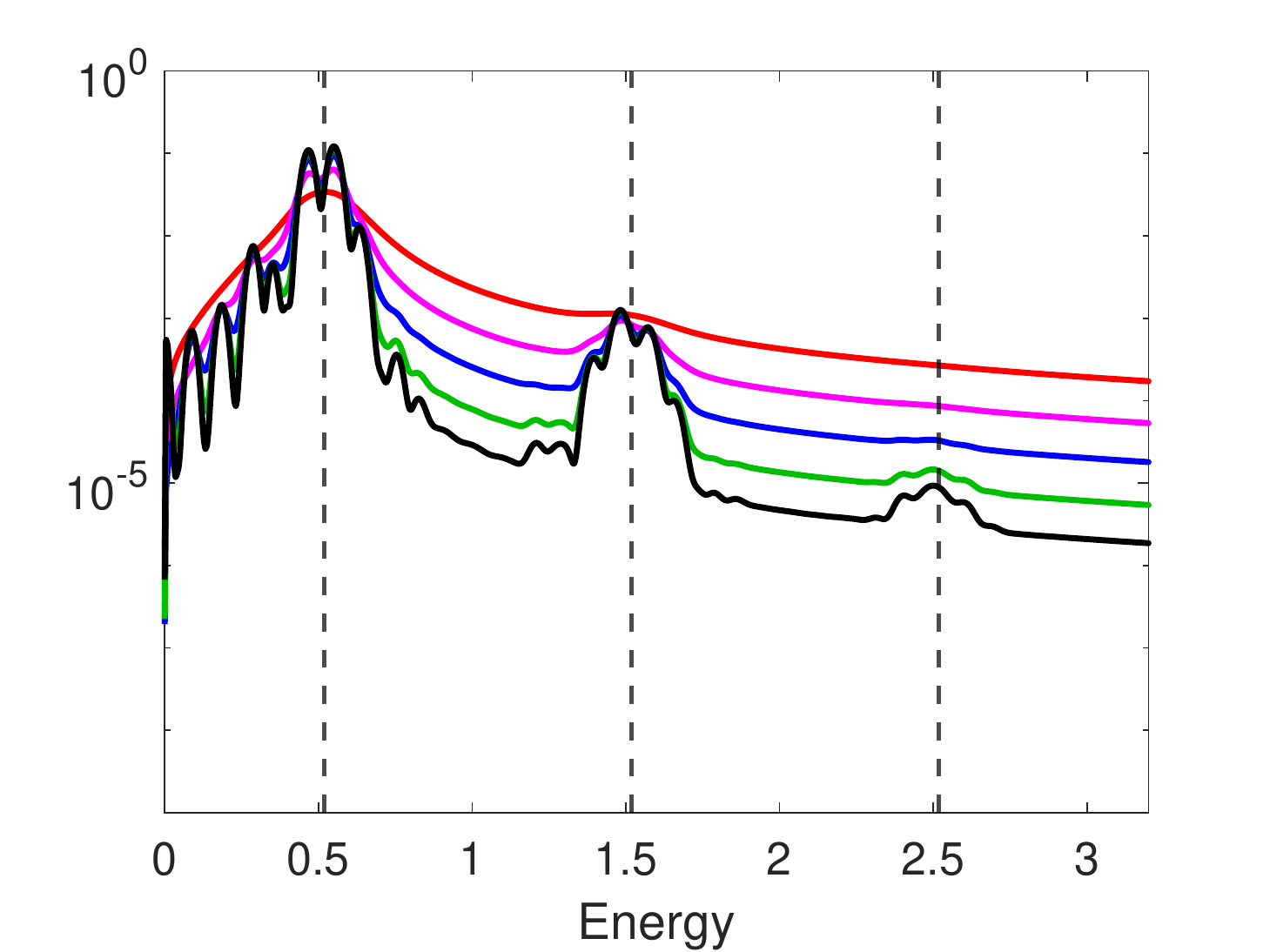}
\end{tabular}
\caption{Upper panel: Energy distribution from the unbound part of the wave function obtained using and energy CAP. The spectra are plotted for various strengths of the energy-dependent CAP function. Lower panel: The energy distributions with a logarithmic $y$-axis. The axes are set in the same manner as in the lower panel of Fig.~\ref{dPdEfromCAPxFigure}.}
\label{dPdEfromEnergyCAPFigure}
\end{figure}

Next we will study the same type of spectra using the energy CAP of Eq.~(\ref{CAPe}). 
We
have chosen an CAP function of form
\begin{equation}\label{RootCAPe}
\mu(\varepsilon) = \mu_0 \sqrt[6]\varepsilon .
\end{equation}
The energy distribution of the liberated and absorbed particle is now provided by Eq.~(\ref{dPdPfromCAPe}), as opposed to Eq.~(\ref{dPdPfromCAPx}) in the preceding case.
The results are displayed for various values of $\mu_0$ in Fig.~\ref{dPdEfromEnergyCAPFigure}. 
I differs from Fig.~\ref{dPdEfromCAPxFigure} in several respects.
One difference is that these spectra are all strictly non-negative -- in accordance with Eq.~(\ref{dPdPfromCAPe}). Another striking difference is how strongly these spectra depend on the CAP strength $\mu_0$.
%
Specifically, ripples are not resolved at all at hard absorption, while the resolution of the structure improves as the absorption strength decreases. The spectra do seem to converge in this limit, albeit quite slowly.

This observation is concordant with the following picture: As the particle enters into the CAP region, it is gradually attenuated and the energy of the absorbed part is recorded as dictated by Eq.~(\ref{dPdPfromCAPe}). Suppose another part of the outgoing wave reaches the energy CAP at a later stage; its energy contribution is, again, recorded and added to the total distribution -- in an {\it incoherent} manner. Thus, these two waves are not allowed to interfere. The absorption is detrimental to any interference patterns which would have emerged otherwise.
Now, this is the same effect as we would see if a {\it detector} where  placed in the CAP region. As the unbound part of the wave function enters the detector region, it would collapse at a certain probability rate. If this collapse is likely to happen before all components of the outgoing wave with the same energy has had the chance to overlap spatially, interference effects will be washed out.
This motivates how an energy-diagonal CAP together with the record of absorbed energy components, Eg.~(\ref{dPdPfromCAPe}), simulates the action of a detector.
The specifics of the CAP function $\mu(\varepsilon)$ would model the characteristics of the particular detector at hand.

In more technical terms: When the measurement consist in projection onto the $|\varphi_\varepsilon \rangle$-basis, the pure state wave function is collapsed according to
\begin{equation}
| \Psi \rangle \langle \Psi |
\label{Collapse}
\rightarrow
\int d \varepsilon \;
\zeta(\varepsilon)
| \varphi_\varepsilon \rangle \langle \varphi_\varepsilon | ,
\end{equation}
where $\zeta(\varepsilon)$ is the distribution function for $\varepsilon$, the outcome of an energy measurement.
The energy distribution of Eq.~(\ref{dPdPfromCAPe}), on the other hand, is the energy-diagonal of an effective, accumulated density matrix:
\begin{align}
\label{AccumulatedDensMat}
& \sigma_\varepsilon = \frac{1}{\hbar} \int_0^\infty dt \; \{\Gamma_\varepsilon, \rho \}
=
\\ \nonumber &
\frac{1}{\hbar} \int_0^\infty dt \; \int d\varepsilon \; \mu(\varepsilon)\{ | \varphi_\varepsilon \rangle \langle \varphi_\varepsilon |, | \Psi \rangle \langle \Psi | \} .
\end{align}
Probability considerations require that the integral of $\zeta(\varepsilon)$ coincides with the trace of $\sigma$; they should both equal the probability of the particle being unbound.
Moreover, if, and only if, the CAP, $\Gamma$, is diagonal in the basis of measurement, the $| \varphi_\varepsilon \rangle$-basis in this case, the diagonal of $\sigma$, $\langle \varphi_\varepsilon| \sigma | \varphi_\varepsilon \rangle = \partial P / \partial \varepsilon $ is manifestly non-negative, cf. Eq.~(\ref{dPdPfromCAPe}). Correspondingly, only then would it make sense to identify the energy distribution obtained with a distribution function such as $\zeta(\varepsilon)$ of Eq.~(\ref{Collapse}).
%

As we have seen, the case is quite different when energy/momentum spectra are calculated using a local CAP, Eq.~(\ref{CAPx}), instead. According to Eq.~(\ref{dPdPfromCAPx}), outgoing waves absorbed and recorded at different times are added together in a {\it coherent} manner; they are still allowed to interfere in momentum space after absorption. Thus, even hard absorption allows interference patterns to be seen, and the emerging spectra turns out to be quite insensitive to the strength of the CAP function.


This suggests that in terms of implementation and simulation, local CAPs are numerically favourable for determining energy spectra as they allow for resolving fine structures while still admitting a strong truncation of the numerical domain. The energy CAP of Eq.~(\ref{CAPe}), on the other hand, has the interesting trait that it simulates the effect of imposing a detector on the system.
Albeit experimental situations usually involve detectors placed far away and with very large extension, corresponding to $R \rightarrow \infty$ and $\mu_0 \rightarrow 0^+$,
we do find the ability to perform such dynamical studies to be an interesting one.
As the
CAP function $\mu(\varepsilon)$ could be virtually any non-negative function,
there is a large degree of flexibility in modelling the detector in this regard.


In the next example we will turn the table and use a local potential, Eq.~(\ref{CAPx}), to model a detector instead.

\subsection{Double slit interference pattern}

\begin{figure}
  \centering
\includegraphics[width=7cm]{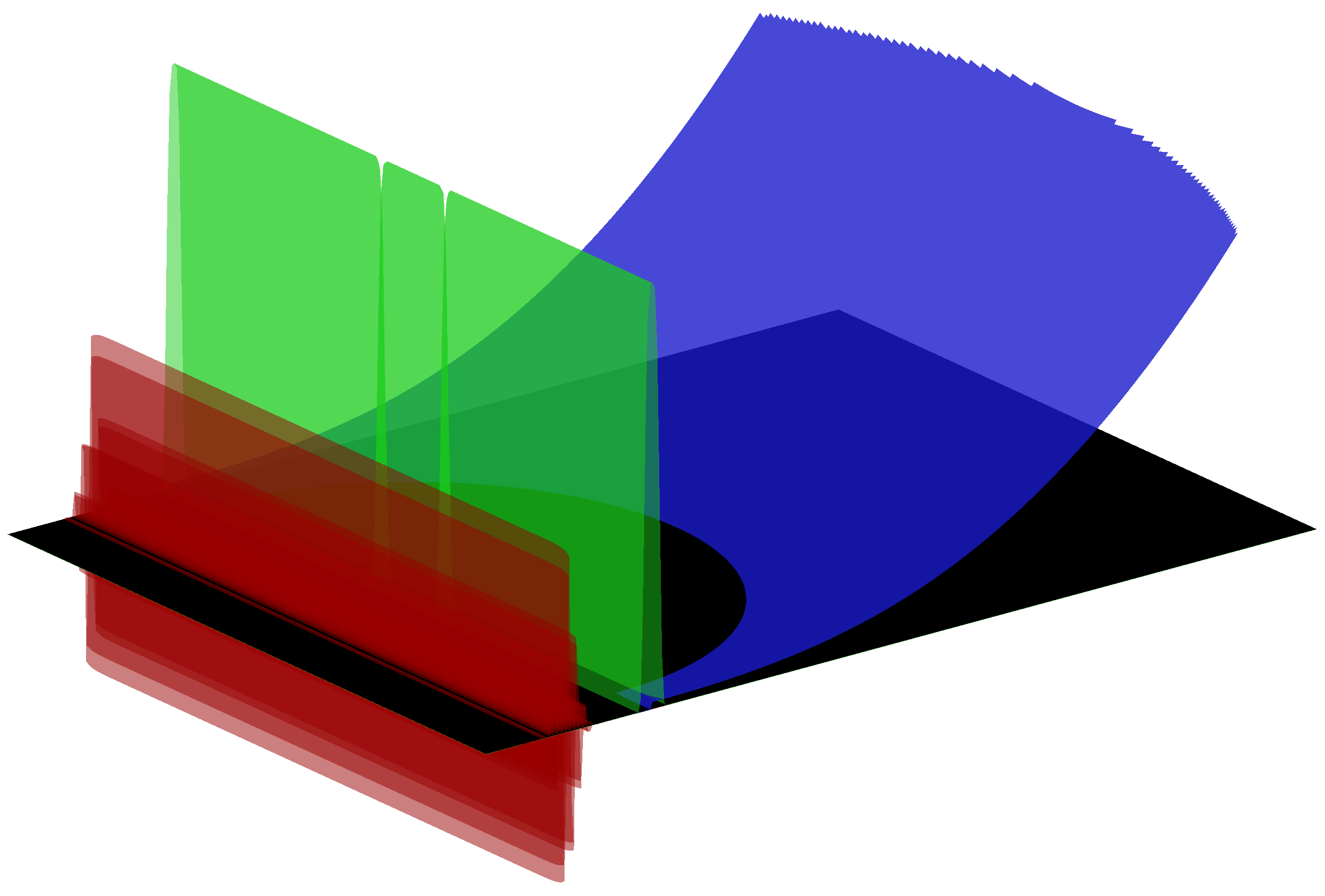}
\caption{The red surface is the real part of the initial wave function, the green surface is the wall/potential -- with two narrow slits, and the blue surface is the absorber/screen.}
\label{DoubleSlitSetupFigure}
\end{figure}

We will study the well known interference pattern emerging from a quantum particle passing through a double slit. The two-dimensional system is illustrated in Fig.~\ref{DoubleSlitSetupFigure}; an initial wave function with narrow spatial extension in the propagation direction travels towards a wall with two narrow slits. Specifically, the initial wave function reads
\begin{subequations}\label{InitialWFdoubleSlit}
\begin{align}
\label{InitialWFpart1}
& \psi(x,y;t=0) = N \psi_x(x) \psi_y(y) \quad \text{with} \\
\label{InitialWFpart2}
& \psi_x(x) = \quad \exp \left(-(x-x_0)^2/(2 \sigma_x^2) + i k x \right) .
\end{align}
\end{subequations}
Here $N$ is a normalization factor, and $\psi_y(y)$ is a near-constant function on the interior of the numerical domain.
The double slit potential is given by
\begin{subequations}\label{DoubleSlitPot}
\begin{align}
\label{DoubleSlitPart1}
& V(x,y) = V_0 \, V_x(x) V_y(y) \quad \text{where} \\
\label{DoubleSlitPart2}
& V_x(x) = e^{-(x/W)^4}, \\
\label{DoubleSlitPart3}
& V_y(y) =
f_\mathrm{FD}\left(y + \frac{d+w}{2}\right) +
f_\mathrm{FD}\left(- y - \frac{d-w}{2}\right)
 \\
\nonumber
&
+ f_\mathrm{FD}\left(y - \frac{d-w}{2}\right) +
f_\mathrm{FD}\left(- y + \frac{d+w}{2}\right) - 1
\quad \text{and} \\
\label{DoubleSlitFD}
& f_\mathrm{FD} (x; T) = \frac{1}{e^{x/T} + 1} ,
\end{align}
\end{subequations}
where $d$ is the separation between the slits and $w$ is their widths. The parameter $T$ adjusts the smoothness of the potential.
At the other side of the wall there is a screen modelled by a local CAP of the same form as in Eq.~(\ref{SquareCAPx}) -- albeit with $x$ replaced by $r$, where $r$ is the distance from the midpoint between the two slits. The situation is illustrated in Fig.~\ref{DoubleSlitSetupFigure}.

We are all well familiar with how waves passing through each of these slits will interfere on the other side of the wall and give rise to an interference pattern. However, this pattern will be altered if a wave emerging from one slit is subject to a position measurement before it has had time to overlap with waves emerging from the other slit. If a screen/detector/absorber is placed extremely close to the double slit and acts over a very narrow spatial region, the corresponding interference pattern will be distorted.

\begin{figure}
\includegraphics[width=8cm]{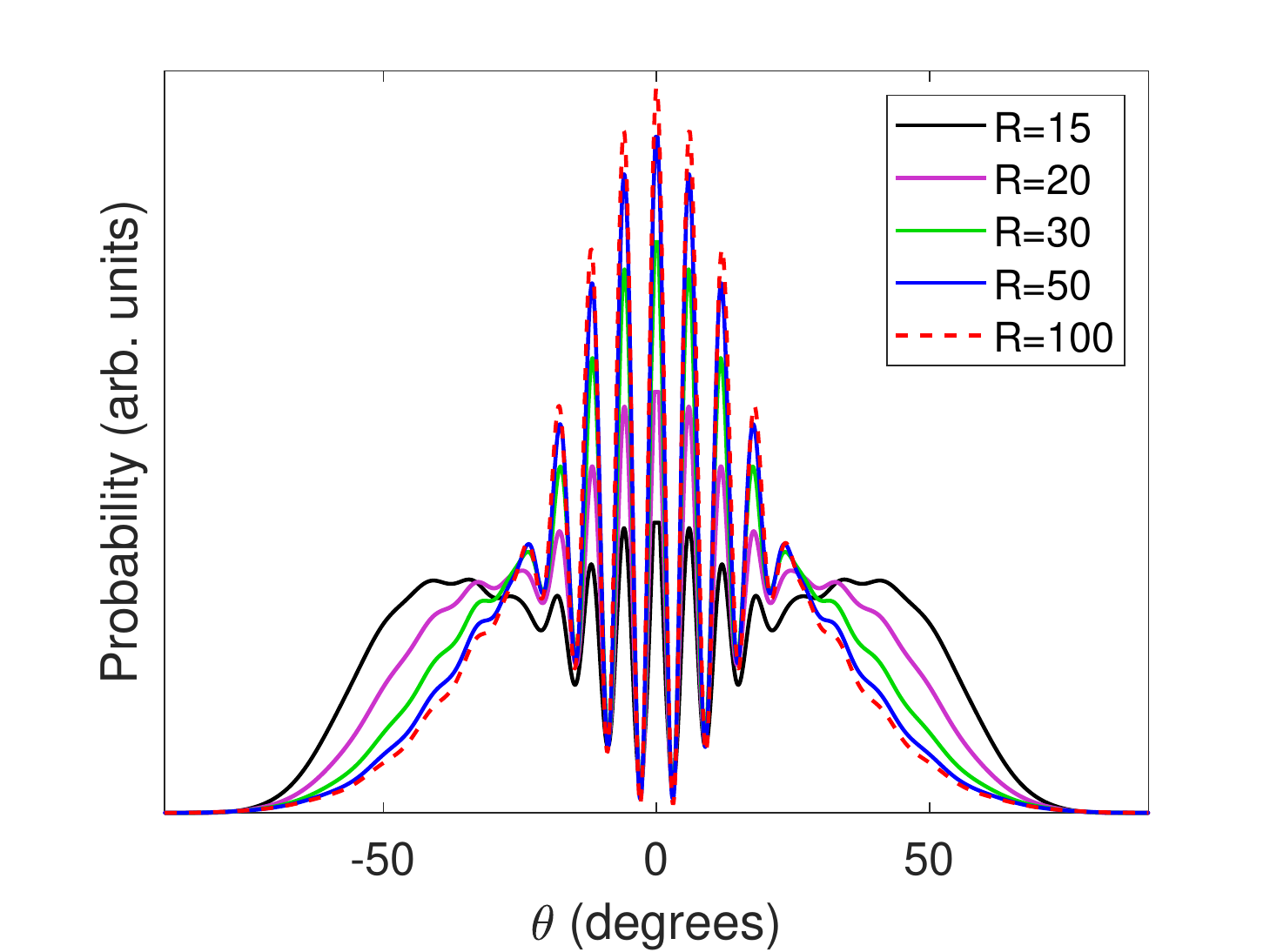}
\caption{The angular distribution on the part of the wave function pertaining to the right hand side of the wall. It is obtained by probing the absorption with absorbers placed at various distances from the double slit.}
\label{AngularDistDoubleSlitFigure}
\end{figure}

Figure~\ref{AngularDistDoubleSlitFigure} demonstrates this for a particular case. In units defined by, again, setting $\hbar$ and the particle mass to one, the initial wave has a mean de Broglie wavelength $\lambda=2 \pi/k$ of $2$. The initial width $\sigma_x$ is equal to $\lambda$, the slits of the potential/wall has the separation $d=20$, centre to centre, and their widths are $w=1.5$. The hight and width of the wall is $V_0 = 100$ and $W = 2$ length units, respectively, and the parameter $T$, which sets the smoothness of the slits, is $0.1$. The CAP potential has the strength $\gamma_0=0.03$~units, cf. Eq.~\ref{CAPx}. We start out by placing the CAP at a distance $R=15$ from the midpoint between the slits and then move it outwards. In each case, the angular distribution is obtained by solving the Schr{\"o}dinger equation with the non-Hermitian Hamiltonian and, in parallel, aggregate the position distribution given by the time-integral in Eq.~(\ref{dPdThetaCAPx}).

As we see, the interference pattern is strongly affected by absorption close to the double slit, while it converges towards a more familiar form as the onset of the CAP is moved outwards. We explain this analogously to the case of Fig.~\ref{dPdEfromEnergyCAPFigure}: From Eq.~(\ref{dPdThetaCAPx}) we see that absorbed waves are accumulated in an incoherent manner. Consequently, if a wave passing through one slit reaches the screen/CAP and is absorbed before it fully overlaps with the waves from the other slit, it will not be  subject to the interference which would have taken place otherwise. The correspondence between this and the action of a position detector is also analogous to what we saw with the energy CAP in the previous example.

\subsection{Concluding remarks.}
\label{RemarksSeq}

Here we have only considered one-particle systems. In \cite{Selsto2021} numerical examples similar to that of Sec.~\ref{Example1Seq} are described, albeit with two particles instead of one -- and only with a local CAP. It is explained how one particle may be absorbed while the remaining particle is still maintained and simulated -- as a one-particle density matrix,
%
The remaining particle may, in turn, go on to be absorbed as well and, thus, also contribute to  energy differential probability distribution of unbound particles. In both cases, i.e., in going from 2 to 1 particle and in going from 1 to 0 particles, the information removed may be aggregated as a an effective one-particle density matrix.
As in Eq.~(\ref{dPdEtillegg}), these density matrices may be analysed by calculating the diagonal elements of interest.
%
%
This also applies to the energy-absorbers, Eq.~(\ref{CAPe}), as also these may expressed as one-particle operators in a multi-particle context.
Thus, the arguments presented here on the close correspondence between CAPs and detectors are not limited to the one-particle case.

In literature, a technique used for analyzing unbound particles referred to as {\it virtual detectors} may be encountered, see, e.g., \cite{Feuerstein2003, Wang2018}. This calls for some disambiguation. Along with a number of similar techniques, see, e.g., \cite{Ermolaev1999, Selsto2005, Tao2012},
it involves the calculation of the probability current, or {\it flux}, through some surface.
While the notion of a detector would, to some extent, seem adequate also in this context, {\it virtual detectors} differ more from
the present framework than the name would suggest. One reason for this is that it does not seem to generalize naturally to the multi-particle case.
Moreover, the CAP is simply used in order to attenuate outgoing waves; it is note used to {\it probe} the outgoing wave. Contrary to the CAP, the {\it virtual detector} does not have any spatial extension, nor  does it bring about any loss in coherence.
%
%
%

The {\it Monte Carlo Wave Packet} approach or the closely related {\it Quantum Jump} method, see, e.g., Ref.~\cite{Dalibard1992,Hegerfeldt1993}, on the other hand, bear strong resemblance to the present approach. Such approaches, along with the derivation of Halliwell~\cite{Halliwell1999}, could hopefully assist in making a closer correspondence between the an actual physical detector and the more generic CAPs used here, be it of the form of Eq.~(\ref{CAPx}), Eq.~(\ref{CAPe}) or any other.
%
%

\section{Conclusions}
\label{ConclusionsSeq}

We have demonstrated how differential information about unbound quantum particles may be calculated using a complex absorbing potential to both attenuate and to probe outgoing waves. When using a local absorbing potential for calculating energy spectra, these spectra are seen to be remarkably robust when it comes to the strength of the absorber. When energy distributions are calculated using an absorber which is diagonal in energy, on the other hand, interference effects are diminished
-- as would be the case if an energy detector is placed in close vicinity of the quantum system.
%

We demonstrated that the same is the case when a local absorber
is used to determine position distributions. Specifically, we showed how an absorber placed close to a double slit setup will prove detrimental to the emerging angular interference pattern.

The close relation between absorption and detection is, thus, underlined by both the fact that information about unbound particles are recorded as the particles are absorbed and that this absorption is detrimental to interference patterns underlines. The latter apply only when quantities in which the absorber is diagonal are recorded. In the non-diagonal case, e.g. when as local absorber is used to calculate energy spectra, coherence is maintained, which, in turn, allows for extracting converged physical spectra on strongly truncated numerical domains.


\end{document}